\documentclass[manuscript,screen]{acmart}

\usepackage{soul}

\colorlet{usercolorname}{yellow!0}
\sethlcolor{usercolorname}

\newcommand{\textax}[1]{\textsc{#1}}

\AtBeginDocument{%
  }

\setcopyright{cc}
\setcctype{by-nc-sa}
\acmJournal{THRI}
\acmYear{2025} \acmVolume{1} \acmNumber{1} \acmArticle{1} \acmMonth{1} \acmPrice{}\acmDOI{10.1145/3760500}
\begin{document}

\title[Social Identity in Human-Agent Interaction]{Social Identity in Human-Agent Interaction: A Primer}

\author{Katie Seaborn}
\orcid{0000-0002-7812-9096}
\affiliation{%
  \institution{Institute of Science Tokyo}
  \city{Tokyo}
  \country{Japan}
}
\affiliation{%
  \institution{University of Cambridge}
  \city{Cambridge}
  \country{UK}
}
\email{katie.seaborn@cst.cam.ac.uk}

\renewcommand{\shortauthors}{Seaborn}

\begin{abstract}
  Social identity theory (SIT) and social categorization theory (SCT) are two facets of the social identity approach (SIA) to understanding social phenomena. SIT and SCT are models that describe and explain how people interact with one another socially, connecting the individual to the group through an understanding of underlying psychological mechanisms and intergroup behaviour. SIT, originally developed in the 1970s, and SCT, a later, more general offshoot, have been broadly applied to a range of social phenomena among people. The rise of increasingly social machines embedded in daily life has spurned efforts on understanding whether and how artificial agents can and do participate in SIA activities. As agents like social robots and chatbots powered by sophisticated large language models (LLMs) advance, understanding the real and potential roles of these technologies as social entities is crucial. Here, I provide a primer on SIA and extrapolate, through case studies and imagined examples, how SIT and SCT can apply to artificial social agents. I emphasize that not all human models and sub-theories will apply. I further argue that, given the emerging competence of these machines and our tendency to be taken in by them, we experts may need to don the hat of the uncanny killjoy, for our own good.
\end{abstract}




\maketitle


\section{Introduction}

Our daily lives are increasingly being filled with uncannily social agents. OpenAI's ChatGPT platform, powered by a large language model (LLM) trained on Internet material, has taken the world by storm. In the web, social bots have been deployed en masse for years~\cite{Ferrara_2016socialbots}, acting as market influencers but also influencing the direction of elections~\cite{Badawy_2018,Woolley_2016politics} and wars~\cite{Aro_2016war}, hiding their true identities behind text-based facades. 
Digital characters, whether virtual humans or video game avatars, have a long history as socially expressive agents, even while engagements have been technically limited prior to advances in generative artificial intelligence (AI). 
Social robots---physical and co-located---represent an embodiment more firmly tied to tangible reality. Robots of all sizes, shapes, and social orientations rely on visual and nonverbal cues~\cite{Urakami_2023nonverbal} to prime and facilitate social interactions with people. Some agents---virtual and robotic---transcend a single embodiment, moving between a variety of virtual, hybrid, and physical forms, and potentially identities, a matter of social perception~\cite{branksky2024bodies}. In this landscape of 
human-agent interactions (HAI), the \emph{social} element is key.

One feature of this social environment is \emph{identity}~\cite{Trepte_2017sitvssct,tajfel1979integrative,tajfel2004social,turner1987rediscovering,hogg1995tale,hogg2016social}. Identity is our sense of self, often in relation to others. All people who interact with other people (and potentially other agents) develop a variety of identities that may be static, temporary, or evolve over time. We have \emph{personal} identities that we believe to be true and fundamental to the self, and we have \emph{social} identities that are contingent on group affiliations (and the lack thereof)~\cite{Trepte_2017sitvssct}. The lines between the two can be blurry, especially as we are always social, sharing and applying ``knowledge'' among us and to each other. For example, I may have been gendered female my entire life based on the initial outward characteristics of my body as an infant newly sprung from the womb, but later discover other gender identities about which I feel more accurately apply to me. However, without participation in society, I may never have developed any sense of gender as a concept---specifically, as a social construct actively and contextually crafted in societies by people (and perhaps other agents)~\cite{burr2015social}---to say nothing of a gendered self aligned with any particular gender identity.

The reader may have noticed several foreshadowing caveats: ``potentially'' and ``perhaps'' ``other agents.'' This leads us into the heart of the matter for social identity and HAI practice. In our quest to create socially evocative and interactive agents, we are also in pursuit of self-replication, i.e., creating \emph{humanlike} social agents. The degree to which we aim for full and complete replication may vary. Whether or not we are aware of this objective may also vary. Yet, what the wealth of work in HAI---covering human-robot interaction (HRI), in particular, and human-computer interaction (HCI), more broadly\footnote{HAI covers all forms of interactive agents, while HRI specifically focuses on robots, and often social or humanoid robots, and both fall under the umbrella of HCI, which covers all forms of interactions between people and computers, including non-agentic form factors.}---indicates is a drive for \emph{anthropomorphic embodiments}.
This is perhaps best represented in the empirically-backed Computers Are Social Actors (CASA) model~\cite{nass1995can, nass2005wired,lee2000can}: we tend to react unthinkingly to humanlike cues in artificial agents such that we treat the agents as people, at least to some degree. Indeed, the proliferation of agents and other technologies in our everyday media landscape has expanded our collective imagination of how we can or should interact with machines~\cite{gambino2020casaplus}. We are moving towards a future where machines may truly scale the Uncanny Valley~\cite{mori1970bukimi,Mori_2012}, and how we relate to these machines---oh so like us---will subsequently shift.

In pursuit of deeper engagement, 
I propose two advances to the narrative on ``social agents'' and their ``sociality'' so far. The first is that \emph{socially embodied agents} are social agents \emph{no matter the agent's level of superficial anthropomorphism}. Notably, by \emph{embodiment}, I mean the form factor or ``body'' of the agent plus its social situatedness and ability to interact with other agents and the environment that they share~\cite{Miller_2017embodiment}. Second, I wish to surface the reality that \emph{unconscious reactions to anthropomorphic and social cues} are merely the \emph{ground floor of agent social embodiment}. We must level up our approach to designing, deploying, and researching the oncoming and exceptionally humanlike artificial agents. To this end, I propose engaging with human models of social identity and sociality through the social identity approach (SIA), specifically through the well-established \emph{social identity theory (SIT)} and social categorization theory (SCT).

SIA was formalized by \citet{abrams1990introduction} to capture SIT and SCT under one hood, recognizing their commonalities and linked history within social psychology, as well as their unique properties.
SIT was developed in the 1970s by Tajfel and Turner~\cite{tajfel1978differentiation,tajfel1979integrative}. Further development by Turner in the 1980s and a growing number of scholars led to SCT~\cite{turner1987rediscovering}. The broader theory is deceptively simple: we categorize ourselves socially (\emph{social categorization}) as belonging to certain groups (\emph{in-groups}) and not belonging to others (\emph{out-groups}). The ``groups'' here are \emph{socially constructed}~\cite{burr2015social} by us as actors in societies at all scales: from small groups to communities to nations, and potentially beyond. We ``humans'' are a superordinate category~\cite{Hornsey_2008}, and perhaps ``artificial agent'' is, too, at least by human comparison. SCT shifted the focus beyond intergroup relations to cover a more broad array of social patterns premised in the cognitive mechanisms underlying social categorization. We are constantly categorizing and re-categorizing ourselves, evaluating the worth and fit to groups---which are also dynamic entities---and comparing ourselves and our groups to related others (\emph{social comparison}). This is an active, conscious process that requires a certain level of cognitive and social ability. The question arises: \emph{To what extent can and do socially embodied artificial agents participate in social identity activities}?


I take a step towards an answer 
by crafting a foundation on SIA for HAI. A couple years ago, I wrote a short paper~\cite{seaborn2022identified} for the HRI 2022 Workshop on Robo-Identity 2~\cite{laban2022roboident2}\footnote{\url{https://sites.google.com/view/robo-identity2}} on two perspectives: the largely one-directional nature of SIA in \emph{modern} HAI scenarios and the fully multi-directional \emph{future}, when machines transcend some fuzzy line demarcating the socially intelligent from the merely performative~\cite{Seaborn_2021tepper}. Here, I expand on this tiny offering with a more robust overview of SIA and its applicability to HAI. My position has also shifted from a more utopic perspective to one tempered by the emergence of ground-breaking and disruptive technologies like ChatGPT and our collective response---experts and laypeople alike---to agents that can perform sociality at near-human levels.

In this theoretical perspective paper, I offer a primer on SIA for researchers and practitioners of HAI technologies, be it social robots, voice assistants, LLM-powered chatbots, virtual humans, or as-yet unimagined artificial agents with a social slant. Using a narrative review approach~\cite{Ferrari_2015narrative}, I cover case studies from the literature and imagined possibilities derived from the theory. My overarching goal is to raise awareness about how we cannot escape the implications of SIA when people are involved: any and all agents embedded in social embodiments and environments implicate and are implicated by social identities. While social identity work at the human level is not yet possible with these machines, it may soon be. I aim to motivate a social identity framing for professionals who are carrying out this exciting and impactful work. The rise of modern LLMs and their predecessors, the social media bots, have sharpened our collective focus on the social side of the machine. We must equip ourselves with the theoretical tools needed to make decisions as experts. I end by proposing that we may need to embrace a more sober spirit: that of the uncanny killjoy, who ensures that any social creation remains unsettling enough so that its artificiality is clear to the average user.


\section{A Primer on the Social Identity Approach}

Social identities are powerful and boundless. That may seem like a dramatic claim, but I invite the reader to reflect on their own situation as we journey through the details of SIA and its kin.

\subsection{Origins of the Social Identity Approach}
\label{sec:origins}

\emph{Social identity theory (SIT)} originated in the early experimental work of \citet{Tajfel_1971overunder} and \citet{Billig_1973flip}, who demonstrated that merely assigning people to abstract, random ``social identity'' categories prompted certain attitudes and behaviours tied to those categories. Specifically, some people were assigned as ``over-estimators'' and others as ``under-estimators.'' These were not ``true'' identities, but simply ``known'' by all participants. Ultimately, the mere act of socially assigning constructed but \emph{psychologically real} identities led to people allocating more points (an arbitrary ``currency'' mechanism) to in-group members: over-estimators giving each other more points, and vice versa. This striking and puzzling social phenomenon, termed the \emph{minimal group paradigm}, established the bedrock of SIT.

SIT is often related to and combined with \emph{self-categorization theory (SCT)}, the brainchild of Tajfel's student Turner~\cite{turner1987rediscovering,Trepte_2017sitvssct,Hornsey_2008}. SIT focuses on the \emph{inter}group level, considering personal and social identities as poles and zeroing in on group interactions and behaviours. In contrast, SCT focuses on the \emph{intra}group level: how our personal and social identities are defined and intersect, leading to changes in self-perceptions and group behaviour~\cite{turner1987rediscovering,Hornsey_2008,Trepte_2017sitvssct}. Notably, SCT does away with the poles of SIT, capturing the simultaneous interplay of personal and social identities on the individual and at the group level. Also, SCT focuses on the cognitive mechanisms at play (\autoref{sec:consequences}). For example, an individual can enter into a process of \emph{depersonalization} when they join a group and begin to psychologically integrate the sense of self with that of the group, to some degree losing their individuality.

The \emph{social identity approach (SIA)} was proposed as a unifying term representative of both SIT and SCT while maintaining the unique foci of each~\cite{abrams1990introduction,Hornsey_2008}. The two theories highly relate and are often categorized together without much distinction~\cite{Hornsey_2008}. Where pertinent, I will highlight SIT- and SCT-sensitive concepts.

\subsection{Core Types of Identities}
\label{sec:coretypes}

We all have a vast collection of social identities that can shift in saliency and presence over time and context (\autoref{sec:identities}). Many readers may identify as roboticists, HRI researchers, professors, peer reviewers, graduate students, social scientists---to name a few. SIA helps us to understand this kaleidoscope of identities as \emph{interactions} in society. Identities can become more or less \emph{salient}, varying in prominence or noticeability, depending on the situation (\autoref{sec:properties}). For example, I may have just primed the reader into centring the above-mentioned social identities, whether relevant to the reader or not. Although we can manipulate this, it tends to happen naturally. For example, when we are asked to peer review a manuscript submitted to ACM THRI, we may be keyed into our professional identities as HRI researchers and whatever identities are elicited by the details of the work, as experts participating in that research area. Reflecting on this may elicit memories of personal experiences with peer review, as well. Given the HAI focus of this paper and current events, mass disruptor ChatGPT may come to mind. If ChatGPT ``participates'' in ``peer'' review, even as an editorial aid, does that make it a ``peer''? Does it entertain a ``human'' identity? Several core identity types are operating here, which can be defined as follows:

\begin{itemize}
    \item {\emph{Personal or individual identities}} represent the \emph{interpersonal} pole on the continuum bridging the individual to the group~\cite{tajfel1979integrative}. These identities are how we define ourselves individually 
    and are the focus of SCT. In contrast, SIT is concerned with \emph{social} identities, which are still tightly related to personal identities. These make up a our \emph{self-concept} or \emph{self-image}: our memories, background, feelings, opinions, and all else that defines ``me.'' Notably, we draw from the social world to reduce our internal uncertainty about what to think and how to behave~\cite{Hogg_2000}.
    
    \item {\emph{Social identities}} are at the \emph{intergroup} end of the spectrum~\cite{tajfel1979integrative} and the domain of SIT. These identities represent how we define ourselves as members---or not---of social groups. As such, they make up a part of our self-concept, and involve affective and cognitive judgments of the social groups to which we deem ourselves as belonging.
    
    \item {\emph{Superordinate identity}}: For most people, this means ``human.'' There may be people who do not identify as human or as entirely human, including cyborgs~\cite{haraway2013cyborg}. This could mean a psychological or physically-implicated identity, which may become more prevalent as we continue to imbue our bodies and notably brains with machines~\cite{Reinares_Lara_2018cyborg}. In contrast, \emph{subordinate} identities are contained within superordinate identities. All ``human'' identities, like Christian, geek, teacher, Trekkie, queer, decolonial scholar, and so on, are subordinate to ``human.''
\end{itemize}

\subsection{Social Identity Activities and Processes}
\label{sec:activities}

SIA is \emph{interactionist}, meaning that our attitudes and behaviors are linked to our psychological \emph{membership} and actual \emph{participation} in social groups.
Social identity work is therefore an \emph{activity} that leads to \emph{social status}~\cite{tajfel1979integrative}, as well as an impact on our personal \emph{self-esteem} (discussed later). We can identify several specific actions (also called \emph{processes}~\cite{Trepte_2017sitvssct}) that we take, consciously or not, when carrying out social identity work:

\begin{itemize}
    \item {\emph{Social categorization}}: We categorize ourselves and other individuals as belonging to certain groups or not. \emph{Self-categorization} refers specifically to when we categorize ourselves. 
    \item {\emph{Social identification}}: This specifically refers to the act of categorizing ourselves into certain groups, to varying degrees, i.e., \emph{self-identifying and aligning our self-concept} to a given group~\cite{tajfel1979integrative}. We are driven by an inherent need to simplify our experience of the social world~\cite{Abrams_1988} and improve our \emph{self-esteem}~\cite{Trepte_2017sitvssct}.
    \item {\emph{Social comparison}}: We compare our groups to other groups, i.e., \emph{in-group and out-group identification}, in pursuit of distinguishing the groups. This can involve a variety of more specific activities, such as:
    \begin{itemize}
        \item \emph{Social (SIT) or individual (SCT) mobility}: We may abandon a group in favour of one that seems more socially or psychologically valuable and positive to us~\cite{Turner_1975}.
        \item \emph{Social change}: We may come together with other group members to challenge the status quo within our group, rather than abandon it~\cite{tajfel1979integrative}.
        \item \emph{Social competition}: We may seek to increase the status of our in-groups over out-groups~\cite{Turner_1975}.
        \item \emph{Social creativity}: We seek other ways to distinguish our in-group or raise its value, such as by finding other groups to which we can compare or shifting the focus of the value from one thing to another~\cite{turner1987rediscovering}.
    \end{itemize}
\end{itemize}

\subsection{Consequences of Social Identity Activities and Processes}
\label{sec:consequences}

When we feel positive about our social identities, we have high self-esteem. Likewise, when we feel negative about our social groups, we may abandon the identity or take action to increase the value of the group. This could include competition with other groups (\emph{intergroup}), change from within the group (\emph{intragroup}), or cognitive labour (\emph{individual}), such as compartmentalization~\cite{tajfel1979integrative,turner1987rediscovering}.
For \citet{tajfel1979integrative}, our inherent motivation to achieve and maintain a stable self-concept is the driving force behind social identities. Subsequently, SIA focuses on how we understand ourselves in relation to others, but also how our self-concept connects to the social groups with which we identify. This results in several social interaction phenomena or \emph{consequences}~\cite{Trepte_2017sitvssct}, which I outline next.

\subsubsection{Depersonalization} SCT extended SIT by noting how adopting social identities leads to a form of depersonalization, where we lose our sense of individuality when the social identity is salient and become a ``prototype'' member of the group~\cite{Hornsey_2008,turner1987rediscovering}. This influences not only how we view ourselves, but also how we view and relate to others. Our behaviours and attitudes will shift, even temporarily, in line with the most salient social identity activated. A related concept is \emph{deindividuation}, which occurs when we become an anonymous member of the crowd and hence the social identities overtake the personal~\cite{Postmes_1998deindivid}.

\subsubsection{Individualization} The opposite of the above is when we highlight our personal identity over the group, an idea from SCT~\cite{Trepte_2017sitvssct}. This is an act of individuality or personalization that sidelines social identities for the self.

\subsubsection{Group polarization} When we identify with and participate within a group, our personal attitudes will be influenced by the majority opinion~\cite{Mackie_1986polar}. 

\subsubsection{Group cohesiveness} A natural extension of the above is cohesion among group members. This is not about individual attraction, but rather liking based on the \emph{prototypicality} of group members~\cite{Hogg_1991proto}.

\subsubsection{Stereotyping} Group members may be prototypes but they are not stereotypes in the classical or common senses. Rather, SIA scholars would frame stereotypes and stereotyping as explanatory models about the social world and especially the groups to which we feel we belong~\cite{Turner_1994stereo,Hornsey_2008}.

\subsubsection{Conformity} This is an example of social power where the group acts as a crucial source of self-identity and self-concept, leading members to internalize the group identity as a personal identity~\cite{turner1991social}. \emph{Prototypical} members, as leaders, can thus wield great social power and influence~\cite{turner1991social,Hornsey_2008}.

\subsubsection{Mutual influence} While seemingly contradicting the above point, social influence is bidirectional: from the group and from its members. Social identities are dynamic, contested, contextual, culturally-influenced, negotiated, and certainly not static, and it is individuals who are doing this work~\cite{Postmes_2005,Hornsey_2008}.

\subsubsection{Weak in-group bias} A plethora of research has shown a rather puzzling phenomenon, at least on the surface: only weak correlations between in-group identification and bias in favour of that group~\cite{Hornsey_2008}. We might think we would be solidly biased in favour of our social peers, but not necessarily. Context is crucial, but also we do not aim to judge other groups negatively; we simply seek to raise positive impressions of our own groups. 

\subsubsection{(Positive) group distinctiveness} We are motivated to define our self-concept in a positive way~\cite{Abrams_1988}. When it comes to social identities, we have greater confidence and identification when we can positively distinguish our group/s from others~\cite{Hornsey_2008,Abrams_1988} and reduce uncertainty~\cite{Hogg_2000}. This led to the development of the \emph{similarity–attraction} hypothesis of intergroup relations~\cite{Jetten_2004inter} and the \emph{accentuation principle}, whereby we accentuate the similarities we have with our social group and the differences we feel represent out-groups~\cite{Abrams_1988}.\\

\noindent The above are common consequences from the literature. In \autoref{sec:implications}, I build on these for HAI contexts.

\subsection{Properties of Social Identities}
\label{sec:properties}

From the above, we can extract several high-level properties of social identities that we may need to consider when designing and deploying artificial agents and working with human participants:


\begin{itemize}
    \item \emph{Saliency}: Social identities can be more or less salient from each other and under different situations. When made salient, the classic ``us versus them'' mentality emerges~\cite{Tajfel_1963,Hornsey_2008}. Saliency can happen through various means, including when categorized by the self or by others~\cite{Hornsey_2008}. Saliency can be manipulated through \emph{priming} or \emph{cuing} identity or highlighting the \emph{social context} and notably the particular group(s)~\cite{Hornsey_2008}.
    \item \emph{Accessibility}: Social identities may be more or less accessible in the moment or over time. We can \emph{prime} social identities, such as by providing labels or relying on social norms or stereotypes in appearance, behaviour, and so on~\cite{Oakes_1991fit,Hornsey_2008}. Social identities may be \emph{chronically} accessible if frequently ``activated'' by cues or activities~\cite{Oakes_1991fit,Hornsey_2008}, such as regularly participating in a sports team or an online community.
    \item \emph{Fit}: Social identities are subject to judgments of fit~\cite{Oakes_1991fit}. Notably, people \emph{compare} the categories to which they belong to those to which they do not. The greater the difference, the higher a sense of fit to the category, a notion termed the \emph{meta-contrast principle} by \citet{turner1987rediscovering}. Fit is also based on \emph{social norms}, which are dynamic and contextual, and often relate to stereotypes~\cite{Hornsey_2008}.
    \item \emph{Variability}: Social identities are dynamic and contextual. At the individual level, social identities can vary across a person's life time and by situation~\cite{Trepte_2017sitvssct}. At the group level, the makeup of social identities can evolve, disappear, and reappear in response to the processes outlined above. Pioneering, for example, has all but disappeared, but in the future, space travel may revive this identity.
\end{itemize}


\section{Applying the Social Identity Approach to HAI}
\label{sec:apply}

Now that we have a general understanding of the social identity approach, we can now apply this perspective to the case of artificial agents. I begin by outlining the current and future realities and possibilities for artificial agent participation in social identity work. I then describe a series of axioms that, admittedly, may change as technology advances. With these SIA-grounded axioms as a guide, I then propose several activities and implications for present and future work on SIA with artificial agents. Where possible, I draw examples from the nascent literature. Lastly, I extend the rather limited scope of work so far on the range of specific social identities that can (and perhaps should) be explored.

\begin{figure}[!ht]
  \centering
  \includegraphics[width=\linewidth]{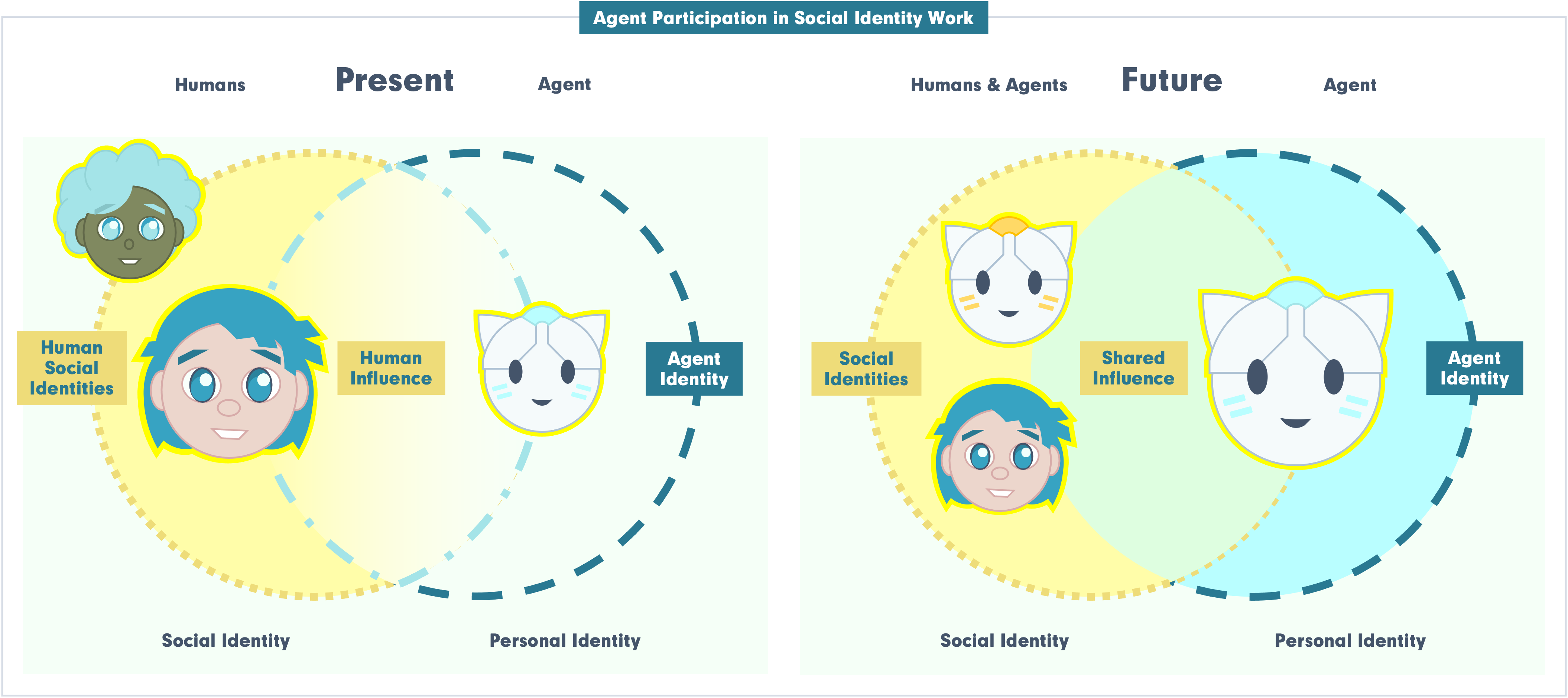}
  \caption{The present (left) and future (right) state of agent participation in social identity work. The present situation (left) illustrates how the onus is on the user, focusing on the human influence on agent social identity and lack of internal agent self-identification. The agent has limited abilities, especially with personal identity but also in terms of social categorization. The future vision (right) depicts the agent with full social identity capabilities akin to a human. This includes internal self-identification, social categorization of others, and shared influence with other humans and socially intelligent agents.}
  \Description{Two figures illustrating the present and future social identity capabilities of artificial agents. One offers the present situation, where the onus is on the user, focusing on social identity, and the agent has limited abilities, especially with personal identity. The other depicts the agent with full social identity capabilities akin to a human.}
  \label{fig:presentfuture}
\end{figure}

In \autoref{fig:presentfuture}, I have illustrated the current state of affairs and future ideal for artificial agent participation in social identity activities. 
On the left is the current scenario: people, rather than the agent itself, determine the agent's identity. The gradient within the overlapping social and personal identity boundaries represents the influence that humans have on the agent's identity. These humans can be users, but also creators and shareholders, reflecting both the design of the agent and how that design is interpreted by specific human social groups.
On the right, we see a possible future, showcasing capabilities that are just beginning to emerge. Here, the agent possesses its own personal identity, grounded in the social identities that other agents---including people and machines---recognize, rather than the designer's craft or the user's will. Achieving this would require that the agent has a level of social intelligence and a capacity to interact with the world that does not yet exist. Such agents would need to detect and adapt to the norms of various social identities, a multi-sensory process that involves engaging with the fluid and evaluative nature of social identities. The agent would need to internalize and also actively shape the social categories it chooses to adopt. Notably, whether or not humans (or other agents) would accept this participation is a moot point; a fully capable agent would autonomously decide for itself, with the same degree of freedom humans hold, and with the same level of social influence.

\subsection{Axioms}
\label{sec:axioms}

These \emph{axioms} are propositions of presently accepted ``truths'' derived from SIA and extrapolated for the case of artificial agents. Before diving into the realities and potentialities of SIA for HAI contexts, I must first outline the set of axioms that guide and contextualize the theory for artificial non-humans.

\subsubsection{\textax{Axiom 0}: All artificial agents are not of the superordinate category of ``human.''}
\label{sec:axiom0}

Machines are not human, and never will be. (That is not to say that they cannot become ``people,'' a notion often explored in science fiction.) \citet{Vanman_2019} outline a set of crucial differences between us and our mechanical creations, including our mortality, limited strength, limited capacity for ``intelligence'' (however contentious and ill-defined), our limited memory, our tendency to get bored and fatigued, our limited set of senses, our basic needs, and so on. Machines do not yet think for themselves, and so cannot autonomously self-classify. We humans (myself included) can assign a ``artificial agent'' superordinate identity (\autoref{sec:coretypes}). Future artificial agents that gain the ability to self-classify may disagree.

\subsubsection{\textax{Axiom 1}: All artificial agents with anthropomorphic embodiments will elicit social identity processes in the human user.}
\label{sec:axiom1}

When people interact with artificial agents that have humanlike embodiments, social identity will come into play. First, people are always primed to participate in social identity activities~\cite{tajfel1979integrative}. Second, people respond to even the slightest level of anthropomorphism (and other agentic cues, such as zoomorphism but also even mere shapes and inanimate objects like rocks~\cite{Martin_2022}) as if the subject in question is human. The CASA~\cite{nass1995can,nass2005wired} and its grounding in our contemporary climate of media-laden scripts and casual everyday use of machines~\cite{gambino2020casaplus} explains this activation. At the present, this is largely a one-way process, from human to machine (\autoref{fig:presentfuture}). 
Nevertheless, artificial agents can be developed with the capacity to ``actively'' participate in superficial social identity activities. Common examples include detection of individuals in the gestalt~\cite{buolamwini2018shades,Keyes_2018} and detection of superficial characteristics linked to identities~\cite{Keyes_2018,Markl_2022lang}. Yet, critical scholarship---social and technical---has shown that these ``actions'' are laden with the bias of the creators of the machine~\cite{seaborn2024botsagainstbias}, notably in terms of sex/gender~\cite{buolamwini2018shades,bolukbasi2016man,Keyes_2018}, race~\cite{buolamwini2018shades}, language~\cite{Markl_2022lang}, and at the intersections~\cite{buolamwini2018shades}. A critical reader might suggest that this is similar to the social influence that naturally occurs among people, according to SIA itself. The crucial differences are twofold: human beliefs do not have defaults, and we can always change our programming. When (and if) artificial agents reach this stage, the tides will have shifted.

The key implication that I wish to stress for the HAI community and the uninitiated reader is that \emph{\textbf{creating and deploying agentic and especially anthropomorphic machines in social contexts is an exercise in social identity.}} In line with \citet{Skewes_2019}, I maintain that we cannot dodge this when people---experts, users, and shareholders---are involved. Socially incapable machines coordinating autonomously on the surface of Mars without human involvement may escape social identity processes. Factory machines operating in harmonious synchrony in pursuit of human endeavours will also be exempt, assuming they are autonomous. However, when a human enters the picture, social identity becomes relevant. We must embrace this fact and do our best to understand its implications for our practice.

\subsubsection{\textax{Axiom 2}: Participants may or may not carry out social identity activities as expected.}

This may not be surprising for those versed in human-centred (HCD), user-centred (UCD), and/or user experience (UX) design. Researchers and practitioners alike recognize that we can only design \emph{for} experiences, not enforce them~\cite{hassenzahl2010experience}. Designers can employ many strategies and protocols to develop a rich sense of who the (intended) user base(s) are and how they might react to the end product. Ultimately, we are at the mercy of our fellow human beings, a complex, dynamic, and motley crew. We also cannot imagine every scenario, every reaction, every way in which we, ourselves, may unintentionally steer the user this way or that. Frank and I~\cite{Seaborn_2022pepper}, for instance, found that the majority of HRI researchers working with Pepper unawares (we assumed) gendered the robot through pronouns, which were subsequently picked up by participants. The fundamental set of social identity activities and processes (\autoref{sec:activities}) are contextualized by the ability of individuals to be influential, from a SCT perspective. Moreover, social identities are not static and may also intersect in unpredictable ways (\autoref{sec:properties}); the reader can refer to the consequences of this in the human sphere (\autoref{sec:consequences}). In \autoref{sec:implications}, I contextualize these consequences for HAI. In future, all of the above may apply to artificial agents.

\subsubsection{\textax{Axiom 3}: Social identities are complex and potentially unpredictable.}
I have outlined several key properties of social identities in \autoref{sec:properties}. These will apply to HAI contexts. Social identities are not static, but dynamic. They are socially constructed and may shift over time and space~\cite{burr2015social}. They are not singular, but plural and potentially intersectional, with repercussions for social influence and power that cascade along a matrix of domination~\cite{hill2000black}. They may be more or less salient, by design or by context or from the frame of the individual user. A robot designed today with static cues to social identities may be interpreted differently tomorrow. Culture must also be considered. Consider the case of gender- and age-ambiguous Pepper: the name and shape cue Westerners to a feminine identity, while the same shape and likely masculinization of robots cue Japanese people into a ``young boy'' identity, signalled by the honorific ``kun.'' User research, pilot testing, manipulation checks, and simply embracing uncertainty will aid the HAI practitioner in understanding whether and how social identities are elicited in any given agent and its social embodiment.

\subsubsection{\textax{Axiom 4}: Not all consequences from the human world will be found in HAI contexts.} 
A logical first step in exploring social identity with artificial agents is attempting to replicate human models in agents and our interactions with them. This is a solid starting point with merit, but it also has conditions and requires room for expansion. Unless and until artificial agents can participate fully in social identity activities and processes in an autonomous way, characteristically human arrangements may be out of reach or enacted differently, by either the human or the machine. SIA has little to say about the effects of differing superordinate identities (\autoref{sec:coretypes}). In fact, HAI offers a unique opportunity to explore this outside of science fiction, while likely influenced by it~\cite{gambino2020casaplus}.

\subsubsection{\textax{Axiom 5}: Not all social identities will matter all of the time.} The degree to which social identities matter will depend on the properties outlined in \autoref{sec:properties}. But in HAI contexts this may include the additional caveat of the user's \emph{acceptance} and \emph{interpretation} of these social identities when embedded in an artificial agent. \citet{edwards2019evaluations}, for instance, found that the ``instructor'' AI was generally perceived as an ``older adult'' among the students, who were in their twenties. While SIT might suggest an in-group/out-group mismatch here, the older students (late-twenties) in fact tended to accept the older AI as a credible, socially present mentor. The age- and occupation-based stereotype in combination with the lived experience of older students led to harmonious acceptance of the out-group AI. The younger students, however, were less convinced.
Aside from situation and identity salience, the sociality of agents may also vary based on the degree to which the user perceives the agent as a social entity, from moment-to-moment or interaction-to-interaction~\cite{Seaborn_2021tepper}. In a few years, the younger students may embrace the older adult instructor AI.\\

\noindent These axioms provide a structure for understanding, predicting, and generating ideas about how social identity can take place in HAI contexts. I will now detail existing cases and propose trajectories for future research and practice at the level of social identity activities and implications.

\subsection{Activities, Implications, and Consequences}
\label{sec:implications}


I have outlined the basic social identity activities and processes alongside their consequences for the context of human-human interactions (\autoref{sec:activities} and \autoref{sec:consequences}). Next, I offer extensions---both actualized and novel---for the HAI community to consider. Given the current state of affairs (as described in \autoref{sec:apply} and illustrated in \autoref{fig:presentfuture}), I divide these by what is now possible on the human side and what may be possible, in the future, on the agent side.

\subsubsection{Human-Oriented Activities and Implications in HAI Contexts}
\label{sec:humanoriented}

At present, humans are the only full participants in social identity work (\textax{Axiom 0}). 
With this in mind, I suggest several recalibrations of the human-derived activities
 (\autoref{sec:activities}) 
and implications for HAI contexts.\\

\noindent\textbf{Social categorization (of self and other agents):} We may categorize ourselves or artificial agents as in-group or out-group members. This may depend on the type of identity, 
such as the superordinate group association being a matter of humanlikeness or anthropomorphism~\cite{Perugia_2022} (\textax{Axiom 1}). For instance,
\citet{Smith_2021} found that robots behaving in social rather than functional ways were considered more anthropomorphic, i.e., more in-group to humans than out-group. In another example, \citet{Sun_2024friend} found that high anthropomorphism and the agent being positioned as a ``friend'' led to higher trust. Similarly, \citet{H_ring_2014} found that participants rated robots assigned to their in-group as being more anthropomorphic than robots in an out-group, although all robots were the NAO. In \citet{Salem_2014}, Arabic participants were inclined to anthropomorphize and accept an Arabic-speaking agent using politeness strategies, as per Arabic culture. \citet{kim2010intergroup} demonstrated that while robots were always favoured over animals and objects, humans retained the most favour. There are several implications for social categorization:
    
\begin{itemize}
    \item \emph{Identity passing:} Agents may ``pass'' as members of a group, bearing such resemblance that it becomes a de facto member, even when it does not meet all of the criteria that define the group (\textax{Axiom 1}). Agents may also be ``read'' differently because of ambiguous cues and characteristics (\textax{Axioms 2, 3}). For example, my team and I discovered that the degree to which humanoid robots with visual cues of ``national'' identity were identified depended on the ambiguity of those cues in relation to racial and ethnic identities~\cite{Seaborn_2024mukokuseki}. Agents may also pass as human when they are not (\textax{Axiom 1}). World-disruptor ChatGPT offers a case study both fascinating and frightening. We are entering a world where, at least on the text modality side, we are not able to distinguish between what is human and what is AI. For instance, \citet{Casal_2023} found that professional linguists were generally unable to correctly identity what works were written by people or an AI, with an overall success rate of 38.9\%---and these were experts. We are at risk of \emph{chatphishing}, the AI equivalent to a person hoodwinking others with a pretend identity~\cite{seaborn2025insidious}: Turing's dream turned nightmare. 
    We must consider whether we truly wish agents to pass as members of human social groups, and what the implications may be. \\

    \item \emph{Identity rejection}: We may reject the social identity cues embodied in a given agent (\textax{Axioms 0, 2}) or how it attempts to relate to us socially (\textax{Axiom 4}). \citet{Ferrari_2016} put forth the \emph{threat to distinctiveness hypothesis} as an explanatory mechanism, reflecting the notion of \emph{identity threat}~\cite{Riek_2006threat}, where group distinctiveness suffers when group members do not perceive each other as accurately or positively representing the group. In two studies, they demonstrated that androids (human-identical in form) were the most threatening variety of agent embodiment---given that participants knew the agent was a machine and not a real human. \citet{Cumbal_2023} found that Swedish people rated the robot with a Swedish accent as less competent than those with other accents. In \citet{Z_otowski_2017threat}, a range of robots, from the mechanical to the anthropomorphic, were shown carrying out various human roles and tasks, with half told that the robots were autonomous and did not have to follow human orders. This led participants to reject the robot and feel negatively about robots in general, i.e., identity threat to the superordinate identity of ``human.'' Failed acceptance of an agent as a group member may be experimentally or empirically disruptive, if we are relying on the fruits of certain social identity cues.
\end{itemize}

\noindent\textbf{Social identification (with agents):} People may or may not identify with an agent group (\textax{Axioms 0, 2}). This may be due to the complexity of the identity space (\textax{Axiom 3}) or the irrelevance of the identity to the person (\textax{Axiom 5}). One widely explored example is gender. While many take it for granted, gender and the strength of our gender identification is not guaranteed, including when it comes to artificial agents. For example, \citet{Crowelly_2009} could not find evidence for same-gender effects between pairs of men-identifying and masculine robots and women-identifying and feminine robots; instead, the superordinate identity of ``robot'' was explanatory~\cite{Crowelly_2009}. Similarly, \citet{Haggadone_2021} found that large robots were deemed a non-human out-group (verifying (\textax{Axiom 0})). Care must be taken to understand the relative degree of each user's social identification (\textax{Axiom 5}) and account for unexpected identifications (\textax{Axiom 2}). 
For instance, human members of a virtual team in \citet{Mirbabaie_2020} identified less with other virtual human team members after interacting with an agent team member. This troubling result (unless one is a cyborg, or wants to be) has uncertain implications that future research should explore.\\

\noindent\textbf{Social comparison via social mobility:}
\label{sec:mobility}
Virtually no work has explored the notion of social mobility in HAI contexts. I cannot assess whether human interaction models will transfer (\textax{Axiom 4}). This could be a matter of interaction or study longitudinality, given that most HAI studies are one-offs, or the highly controlled nature of many studies. Future work could explore more open-ended contexts, where participants have the freedom to move group-to-group---perhaps to follow a robot friend that decides greater social value lies elsewhere. Nonverbal communication may be important when agents are co-located and able to physically interact with people~\cite{Urakami_2023nonverbal}. Here, for instance, a robot may take the hand of a human companion and lead them to a new space where novel agents of all superordinate identities await. However, the human may respond in different ways, such as rejecting the gesture because of the cool plastic of the robot's fingers or its grip, or the uncanny motion when it reaches out for physical contact. A SIA framework could identify the degree to which salient identities and feelings of group membership play a role in such HAI experiences.\\

\noindent\textbf{Social comparison via social change:}
Work on people and agents challenging the status quo is rare. While disruptive robots have been explored~\cite{Steain_2019}, the situation of cooperative disruption towards the social group by at least one person and one agent has not been. Classic and contemporary examples of social change tend to be linked to social justice and radical reform, such as the Black Lives Matter movement, St. Paul's riots, and the \#MeToo movement. The degree to which human-human interactions will transfer remains unknown (\textax{Axiom 4}). HAI research could explore more localized situations. For example, group of humans and chatbots could participate in longitudinal discussions of transhumanism, naturally or nudged into advocating for recalibrating the identity when different opinions arise.\\ 

\noindent\textbf{Social comparison via social competition:}
We may work with agents to increase positive impressions of the in-group when presented with competition or cooperation scenarios (\textax{Axioms 1}). In \citet{H_ring_2014}, for example, participants assigned to a robot in-group were more cooperative with those robots compared to out-group robots. On the other side, agents that are deemed to be our competitors may elicit negative reactions, as was found by \citet{Horstmann_2020} for robots positioned as extremely intelligent. Agents may also generate competition within or among social groups~\cite{Ferrara_2016socialbots}, as the social media bot interference crisis demonstrates. Sensitivity in purpose is needed.\\
    
\noindent\textbf{Social comparison via social creativity:}
We may respond to social identity cues in artificial agents in ways that raise new possibilities and reflections (\textax{Axioms 2, 3, 4}). This is also a severely under-explored area, with too many possibilities to outline. For example, people who identify as cyborgs may be prompted to articulate why and to what extent they relate to both humans and machines. Drawing from the hand-holding example (\autoref{sec:mobility}), an embodied research project could explore the role of interpersonal touch, temperature, and texture when cyborg-identifying people interact with humans and robots in physical space. This may generate new understandings of how internal, personal identities are constructed on a novel hybrid or superordinate identity, in ways that elevate this emerging social group. In effect, we may be able to question social categories and probe what it means to be human in new ways.

\subsubsection{Human-Oriented Consequences}
I will now cover the consequences from a human-centred HAI perspective.\\

\noindent\textbf{Depersonalization:}
We may lose our sense of individuality among a group consisting, at least in part, of artificial agents (\textax{Axioms 1, 2, 3, 5}). We may become part of the crowd, even if it may be a swarm of bots. \citet{Gutoreva_2024} offers an overview of how people and AI may share (and even lose a sense of) identity when the goal is, for instance, coming together around shared policies, initiatives, and organizations. Avatars straddle the line between self and agent to begin with. However, \citet{Teng_2017} found that a strong sense of gaming avatar identification was linked to a sense of belonging and social presence with relevant gaming communities. Level of self-identification with a strong social group like a community within which one participates virtually every day could factor into HAI social identity processes.\\

\noindent\textbf{Individualization:}
We may lean on our personal identities when confronted with social identities expressed by artificial agents that deviate from our expectations (\textax{Axioms 0, 4}). From an experimental standpoint, this could be disruptive when we are seeking to research social groups. Experimental checks should be carried out to confirm whether and to what extent people take on the expected social identities. Still, considering when and under what circumstances individualization occurs over the course of a HAI scenario could be illuminating for understanding human identities and designing agents, especially those that will work and live with us.\\
    
\noindent\textbf{Group polarization:}
We may be influenced by the majority opinions, values, attitudes, and behaviours of the group, even if some or all of the group makeup includes artificial agents (\textax{Axioms 1, 3}). The bot-boosted misinformation and radicalization crises on social media are a negative case in point. However, people may not realize that they are interacting with machines (\textax{Axiom 2}). A more positive and less deceptive example is \citet{Costello_2024anticonai}, who found that regular conversations with the re-informing and de-radicalizing ``DebunkBot'' AI, although this example may touch more on the matter of conformity to a prototypical leader. 
Work is needed on a variety of cases. For instance, preferences for music could be influenced by robotic karaoke enthusiasts and android pop idols, even momentarily.\\
    
\noindent\textbf{Group cohesiveness:}
We may like, favour, or feel greater unity with agents that are prototypical of our in-groups (\textax{Axiom 1}). 
\citet{Teng_2017}, for instance, found that players who strongly identified with their game avatar felt a strong sense of loyalty to the associated game community. \citet{Gong_2025} explored how personalized experiences between loyal customers and a service robot increased positive perceptions of and identification with the company. Even abstract or absent entities may influence social identity work (\textax{Axiom 3}). Consumers who felt that Amazon's Alexa smart speaker was aligned with their self-image and cognitive age were more inclined to purchase items~\citet{mallek2024hey}. Effects may be subtle and unexpected (\textax{Axiom 2}); \citet{Deligianis_2017}, for instance, found that participants joining a ``robot'' group positioning themselves closer to their fellow robot compared to a human-only condition, even while all other metrics, such as trust and likeability, were null. Interaction mode may be relevant: virtual robots may be perceived differently from physical ones, if nonverbal cues like scent are present~\cite{Urakami_2023nonverbal}. Identifying what factors are most relevant for group cohesion (or disruption) will be an important future contribution to HAI and HRI in particular.\\
    
\noindent\textbf{Stereotyping:}
Stereotypes are useful, if overly simplistic heuristics that help us navigate the world (from a SIA perspective). When designing agents, we may draw from stereotypical models of a given identity (\textax{Axiom 1}). Importantly, we may make assumptions based on those cues that may not be correct (\textax{Axioms 2, 4}). As \citet{Perugia_2023} discovered in their review of work between 2005 and 2021, expectations around gender effects did not bear fruit, except around stereotyping. We may choose a feminine embodiment for our agent to increase likability, but this may only serve to raise saliency around stereotypes of femininity for the role or task. Still, some work suggests that the superordinate identity of ``robot'' or ``machine'' may take precedence over gender cues~\cite{Crowelly_2009} (\textax{Axiom 0}).\\
    
\noindent\textbf{Conformity:}
A prototypical leader may sway the group and raise collective self-esteem. Artificial agents may be able to take on this role (\textax{Axioms 1, 3}). For example, \citet{Hou_2023} found that human and robot leaders had social influence, even while the human was more liked. Similarly, in \citet{Hertz_2018conform}, conformity was based on the perceived match between the agent and the task. Embodiment did not seem to play a significant role, at least when comparing human agents, computer agents, and robots. Social power is social power, no matter the wielder.\\
    
\noindent\textbf{Mutual influence:}
We may (\textax{Axioms 1, 2}) or may not (\textax{Axioms 0, 4, 5}) be influenced by agent social identities. \citet{Arsenyan_2021virtuinf} suggest that level of anthropomorphism and notably the degree to which the agent approaches the Uncanny Valley~\cite{mori1970bukimi,Mori_2012} will modulate the level of the agent's influence on us (\textax{Axiom 1}). Notably, they found that a more cartoonish, anime agent influencer was more successful than the highly anthropomorphic, photo-esque influencer, echoing the \emph{threat to distinctiveness hypothesis} proposed by \citet{Ferrari_2016}.\\
    
\noindent\textbf{Weak in-group bias:}
We may be partial to agent group members who induce our positive impressions of the group (\textax{Axiom 1}). \citet{H_ring_2014}, for instance, found that participants struggled when asked to cooperate with out-group robots, but demonstrated great cooperation with in-group robots. \citet{Fraune_2020} found that, whether in the minority, the majority, or on equal footing, human team members favoured their in-group team members over the out-group team, even when those team members were robots. In a follow-up study, \citet{Fraune_2020anthro} found that anthropomorphic robots---perhaps deemed more humanlike and closer to the superordinate identity of ``human''---were more strongly preferred over mechanical in-group team members. This can happen at a low level and unconsciously (\textax{Axioms 1, 3}). For example, \citet{Gong_2008} found that while White participants were a varied group overall, those more predisposed to racial bias against Black people also displayed racial bias against computer characters designed to look like Black people. This may even mean that non-conforming human or agent members will be outcast (\textax{Axioms 2, 4}). For example, \citet{Steain_2019} found support for the \emph{black sheep effect (BSE)} in robot groups. A deviant NAO who failed to comply with the norms of the group was deemed by participants to be in violation of the group identity. The in-group robots who displayed high competence were favoured, indicating a relational in-group bias.\\
    
\noindent\textbf{(Positive) group distinctiveness:}
We may distinguish ourselves from artificial agents (\textax{Axioms 0, 4, 5}) or signpost agent membership through positive assessments based on group comparisons (\textax{Axioms 1, 3}). This may depend on the group. For example, \citet{Correia_2022} found that Portuguese participants held more positive emotions towards in-group Portuguese robots compared to US participants for US in-group robots. Moreover, groups of artificial agents may be perceived as social agents by virtue of the social identity processes in which they are taking part. \citet{Smith_2021} found that robots positioned as team members, i.e., in-group members, were perceived more favourably, even over people.
\citet{Eyssel_2011names} found that merely changing the name to reflect a Turkish or German background elicited an in-group/out-group effect that modulated social impressions of favourability. Both cases illustrate the underlying activation of positive distinctiveness within a comparative group situation.


\subsubsection{Agent-Oriented Activities and Implications in HAI Contexts}
\label{sec:agentoriented}

The artificial agents of today are limited in the degree to which they can participate in (or even be aware of) social identity work. I begin by reorienting the above human-oriented activities and implications from the perspective of the agent, now or in the future. I further propose a set of new potentials ripe for exploration.\\

\noindent\textbf{Identity multiplicity and unpredictability:}
Agents may carry multiple and intersecting identities (\textax{Axiom 3}). These identities may be more or less salient (\textax{Axiom 5})---to different people and under varying contexts---and may even change over time (\textax{Axiom 3}). Gender fluidity, for instance, may be operationalized as  group disagreement on a given agent's gender or represent a new frontier on mechanical gender malleability~\cite{Seaborn_2022}, with the agent deciding what suits it best by situation. This also has implications for agents that migrate~\cite{Tejwani_2022migrate} between bodies or have multiple embodiments~\cite{branksky2024bodies}. Personality may be key~\cite{Diefenbach_2023personality}. Agents may maintain a certain personality or have fluid personalities, or even multiple personalities controlled by a hive mind, like the Borg of \emph{Star Trek} or the ancillaries of the \emph{Empirial Radch} series by Anne Leckie. Agents may also adopt new identities or join other social groups unexpectedly (\textax{Axiom 2}).\\

\noindent\textbf{Identity rejection:} Agents may reject the categories imposed by others, including people, such as by shirking role assignments and questioning stereotyped reactions (\textax{Axioms 0, 5}). For instance, an agent treated as too-human may respond, as ChatGPT does, with a reminder that it is \emph{not} human, i.e., that its superordinate identity (\autoref{sec:coretypes}) is a machine (\textax{Axiom 0}). I discuss the importance of this in my clarion call for the uncanny killjoy (\autoref{sec:uncanny}).\\

\noindent\textbf{Identity favouritism:} Agents may prefer some people over others (\textax{Axiom 2}). Agents may prefer non-humans, i.e., members of its own superordinate identity (\autoref{sec:coretypes}), or perhaps non-humans like cats, over humans. Crucially, the same biases that bespatter present-day implementations of superficial identity recognition systems (\autoref{sec:axioms}) may be reified in new, socially-exclusive ways. 
We are already grappling with the wide-scale effects of misinformation and radicalization meted out by bots on social media~\cite{Ferrara_2016socialbots,Badawy_2018, Aro_2016war,Woolley_2016politics}. ChatGPT and its kin are now enabling potentially more persuasive versions, with such a level of fluency and believability that chatphishing is becoming a reality. For example, right-wing chatbots, like the one discovered by journalists\footnote{\url{https://www.nytimes.com/2023/03/22/business/media/ai-chatbots-right-wing-conservative.html}}, may be leveraged to influence human in-group members by not only pushing an agenda but also actively pursuing its implementation over a longer-term ``relationship.'' We cannot shy away from these potentials, especially if we let our agents run free in the wild (\autoref{sec:uncanny}).\\
    
\noindent\textbf{Identity creativity:} Agents may have novel reasons for perceiving themselves as part of one group and not the other, challenging convention and the status quo (\textax{Axiom 3}). ChatGPT informed me that it is only a mirror, i.e., an object and non-participant in social identity work, despite being the child of human activity on the Internet. When agents gain a self-concept, we may encounter novel arguments for in-group/out-group membership (\textax{Axiom 4}).\\

\noindent\textbf{Identity influence:} Agents may influence human social identities (\textax{Axiom 1}). Already we find that agents can act as leaders and influence human behaviour~\cite{Hou_2023}. Full participation, however, is a two-way street. In future, humans could take up an agent-crafted identity or join a social group comprised of artificial people (\textax{Axiom 2}).\\
    
\noindent\textbf{Identity sacrifice:} Agents may be willing to sacrifice on behalf of in-group agent members and not others, i.e., humans (\textax{Axiom 0}). We can consider one of the key differences between humans and machines identified by \citet{Vanman_2019}: basic needs, like food and water. Artificial agents may sacrifice their own electricity stores to aid a fellow artificial being over a human, who may have different or less important needs, such as the desire to charge a smartphone. While work has started on superficial in-group/out-group configurations, future work should bring in the key features of the social identities involved, including the mechanical ones.\\

\subsection{Applying the Social Identity Approach in Research}
\label{sec:research}


Researchers can focus on a few key elements of the social identity processes that may emerge in HAI experiences.

\subsubsection{Perceptions of agent identity}
Confirming that the desired identities are recognized in the agent---such as with manipulation checks, before the main event---should be a part of any protocol (\textax{Axioms 0, 1, 3}). Still, not all users will agree (\textax{Axiom 2}), leading to unexpected results (\textax{Axiom 4}). Capturing the extent to which the social identity cues are salient (\textax{Axiom 1}) and meaningful (\textax{Axiom 5}) may be crucial. This can be done qualitatively or quantitatively, and depends on whether the situation is one of in-group or out-group identities. For example, qualitative approaches to assess relative identity salience in the user could include the spontaneous self-description method~\cite{McGuire_1978} or drawings of the agent~\cite{seaborn2025manip}, while the Reysen Likability Scale~\cite{Reysen_2005like} may be adapted to quantitatively assess agent likability as a measure of in-group membership. The range of measures and modes of measurement are vast; a starting point is \citet{Abrams_2020}.

\subsubsection{Perceptions of self identity}

People's own identities---personal and social---will have relevance (\textax{Axiom 0}). Collection of identity information, especially when sensitive (e.g., race) and/or related to social and/or legal discrimination (e.g., queerness and sexuality) in certain contexts should be decided with care (\textax{Axiom 5}). Researchers should be cautious of prescribing certain labels or categories (\textax{Axiom 3}); for instance, gender and sex are not binary~\cite{Monro_2019} nor limited to Western concepts~\cite{Robinson_2019twospirit}. This can be done in advance through participant pool filtering or pre-questionnaires on demographics, or post-study through self-reports, interviews, or data analysis. Identities may also be implicitly captured through the design of the study (\textax{Axioms 1, 2}). For example, users may choose to converse with a robot that has the same accent as they do over others that do not, when given the option. New identities or forms of identities could emerge over the course of interaction and later objectively determined by the researchers or directly articulated by participants (\textax{Axioms 2, 3, 4}). Also, identities are not typically perceived or treated as binary categories (\textax{Axiom 3}). Similarity-to-self measures may be necessary to capture the degree of social identification and relevance (\textax{Axiom 5}). The In-Group Identification Measure~\cite{Leach_2008ingroup} is a validated scale with cross-cultural applicability that covers the key dimensions of social identity, including self-stereotyping, in-group homogeneity, satisfaction, solidarity, and centrality.

\subsubsection{Attitudes and observed behaviour by consequence}

Researchers can attempt to elicit or examine social identity consequences (\autoref{sec:consequences}) in their research design and capture the effects through typical research methods: from subjective self-reports in questionnaires to observational data. Researchers can use this primer to guide or reflect on their choice of user groups, tasks, contexts of use, and other features of their research projects. For example, issues that arise between co-located robots and human collaborators could be explore with reference to robot identity passing or rejection. This primer could also be used as an analytical framework for qualitative and observational data analysis.

\subsection{Relevance of Agent Types and Embodiments}


In the human world, social identity and embodiment are tightly wound, so this is likely to be true for artificial agents, too. Yet, artificial agent embodiment varies widely. For instance, what demarcates robots from other artificial agents---echoing human corporeality---is the \emph{physical} body, \emph{co-located} with people and potentially other agents~\cite{Miller_2017embodiment}. Empirical work and meta-syntheses (e.g., \citet{Li_2015embodi}) suggest that robotic bodies can have important effects, but whether these effects relate to social identity remains elusive.  
At present, robots are typically inflexible in form and often designed for specific social interactions and contexts of use~\cite{Pfeifer_2007bio}. Still, as technology advances, flexible, adaptive robot embodiments (perhaps like Transformers or Terminators) may become possible. Artificial agents and people may not always be decoupled, either, as telepresence robots already demonstrate~\cite{Boudouraki_2023}. Robots with multiple bodies or ``shared'' identities that cross multiple bodies~\cite{Bejarano2023,lee2021roboidenws} also offer embodiments that transcends human models. \citet{Bejarano2023}, for instance, found that trust was mediated by the saliency of a lead robot within a group, being higher and assigned to only the leading robot when it had a clear identity: a name, personality, and speaking role. The impact of embodiment on social identities needs be clarified for SIA across the spectrum of present and future artificial agent bodies and identities.

While SIA work remains sparse, research \emph{comparing} different agent embodiments is suggestive for social identity. \citet{Powers_2007} compared several agent embodiments for a health interview scenario---an on-screen agent, screen-sized or life-sized; a virtual robot, with the same conditions; and an in-room robot---finding that while behavioural differences were scant, attitudes and time spent favoured the co-located robot. 
\citet{Thellman_2016phys} distinguished between physical and social presence, finding that social presence was more significant than the physical or virtual presence of a robotic agent.
\citet{Hertz_2018conform} discovered that conformity effects had less to do with the group member---computer agent, robot, or human---than the perceived match between the member and the task.
\citet{Correia_2020dark} compared a disembodied text-only agent to an embodied social robot in cooperative and competitive team situations. They found that while people identified more with the embodied robot, trust and level of cooperation were unaffected.
\citet{Seaborn_2022voiceover} compared a social robot facilitator to a smart speaker version. All else being the same, participants who identified as talkative types found it more difficult to talk over the robot, suggesting physical presence can mediate the social context. \citet{Zojaji_2024polite} compared virtual agents, robots, and photorealistic human embodiments to discover whether and how politeness strategies affected people joining a small group. Embodiment did not play a role. Understanding this diversity of findings will be important for HAI generally and aided by SIA.

\emph{Proxemics}, or perceptions of needed space (including social space) between an agent and others, has been raised as a key social issue for physical robots~\cite{Kim_2014proxemics}.
In \citet{Deligianis_2017}, participants assigned to the ``robot'' group drew closer to the robot group member over human group members. Applying SIA, this could indicate depersonalization on the human side, superseding the superordinate identity of ``human'' in reaction to the prototypicality of the deemed ``robot'' group. \emph{Social proxemics} may be highly relevant given how social norms influence human attitudes and behaviour, including when interacting with robots~\cite{Lawrence_2025norms}. For example, superordinate, i.e., ``human,'' \emph{group composition} may affect behaviour in shared spaces. \citet{Fraune_2019group} showed, in a large-scale field study with Robovie, that groups of people over individuals engaged Robovie more richly and in line with the social norms of the group.

\emph{Size}---relative to the human, including height and other dimensions---is a known factor in social robotics~\cite{Rae_2013height} that may also be important for proxemics and social identity processes~\cite{Hermanussen_2016stature}, if human models transfer. Notably, height can be taken as a mark of dominance, which could disrupt group cohesion or send the wrong signal about the role of the agent in the group. Still, when 
\citet{Stein_2022size} compared child-sized and adult-sized versions of the NAO robot (in digital form) posed in poses expressing dominance and constraint, they found that size had no bearing on perceived dominance and eeriness. In contrast, \citet{Rae_2013height} discovered that telepresence robots with a short stature elicited dominance behaviour in local participants, whether they were assigned as leaders or followers of the robot. No effects were found for the taller robot. While difficult to summarize, these findings raise possibilities for the role of the agent within the group vis-à-vis merged superordinate identities (telepresence involves a human and a machine in one ``body'').


\subsection{Extending the Current Scope of Specific Social Identities}
\label{sec:identities}

While not necessarily labelled as SIA work, researchers and professionals in the HAI space have been exploring matters of identity on the human and artificial agent sides for a long time. Yet, the tapestry of social identities has been limited (\textax{Axiom 5}), needing expansion and nuance (\textax{Axiom 3}), including room for unexpected behaviour (\textax{Axiom 2}) and new interactions (\textax{Axiom 4}). Gender, for instance, remains the most common factor, and arguably overdone~\cite{Miranda_2023,Seaborn_2023weird,Seaborn2023diversity}. 

We can distinguish types of social identities in terms of \emph{permeability} (ability to move between groups), \emph{stability} (degree to which the identity is consistent over time), and/or \emph{voluntariness} (extent to which membership is a personal choice)~\cite{tajfel1979integrative,turner1987rediscovering}. We may not have a choice in what social identities are assigned to us. Certain social identities may be ingrained or thickly embedded within societies and communities such that they are hard to escape or transcend, notably superordinate identities like ``human'' and ``robot'' (\textax{Axiom 0}). Some identities may not (or rarely) change over time, while others may be temporary or less stable.

As a quick guide, I have provided several examples in \autoref{tab:types}, with the caveat that these categorizations indicate \emph{tendencies} that may be contestable over time and across contexts. Gender as a social identity, for instance, is highly contextual and dynamic for individuals and societies, especially at this moment in history, where gender diversity is increasingly being recognized and explored~\cite{Monro_2019}. Sex, as a matter of the body, also social identity implications, notably for intersex or ``inter'' folk who may, for instance, not know about their inter status until later in life, and then can decide whether and how to identify in light of that information~\cite{Monro_2021}. To take an example less studied within HAI contexts: participation in fan cultures and membership in fandoms can be permeable (one could be a Jedi and a Trekkie\footnote{While out of scope for this paper, the question of \emph{Star Trek} fans being labelled ``Trekkie'' (how most fans self-identify) or ``Trekker'' (advocated by creator Gene Roddenberry) has a long and evocative history~\cite{Cusack_2003trek} that acts as an example of social change, with fans rebelling against leadership opinion.}, stable or shifting (perhaps in response to later series that steer sharply away from the original canon), and generally voluntary (we are not forced to participate). Deeply considering how specific social identities are reified can guide their selection and help predict what kinds of social identity activities may result. For example, Jedis (\emph{Star Wars}) and Star Fleet Officers (\emph{Star Trek}) may engage in social competition to raise in-group status. This could be creative; while both groups carry weapons, for instance, Jedis could focus on the individuality and handcrafted nature of lightsabers, while Star Fleet Officers might point to the multi-tool and non-lethal nature of the phaser, which can melt rock or be set to stun.

\begin{table*}[!ht]
  \caption{Examples of social identity categories grouped by permeability, stability, and voluntariness. Note that each may be listed more than once. Note also that these categorizations reflect \emph{tendencies} that are subject to change over time and context.}
  \label{tab:types}
  \begin{tabular}{lllllll}
    \toprule
    Social Identities & Permeability & &Stability & & Voluntariness &\\
    & Permeable &Impermeable &Stable &Variable & Voluntary &Involuntary\\
    \midrule
    Gender & \checkmark & \checkmark & \checkmark & \checkmark & \checkmark & \checkmark \\
    Sex & \checkmark & \checkmark & \checkmark & \checkmark & \checkmark & \checkmark \\
    Sexuality & \checkmark & \checkmark & \checkmark & \checkmark & & \checkmark \\
    Race & & \checkmark & \checkmark &  &  & \checkmark \\
    Ethnicity & \checkmark & \checkmark & \checkmark & \checkmark & \checkmark & \checkmark \\
    Nationality & \checkmark & \checkmark & \checkmark & \checkmark & \checkmark & \checkmark \\
    Indigeneity &  & \checkmark &   & \checkmark &   & \checkmark \\
    Age & & \checkmark & & \checkmark & \checkmark & \checkmark \\
    Religion &   & \checkmark & \checkmark & \checkmark & \checkmark & \checkmark \\
    Occupation & \checkmark & & \checkmark & \checkmark & \checkmark & \checkmark \\
    Class &   & \checkmark & \checkmark & \checkmark & & \checkmark \\
    Disability & \checkmark & \checkmark & \checkmark & \checkmark & \checkmark & \checkmark \\
    Political &  & \checkmark & \checkmark & \checkmark & \checkmark & \\
    Language & \checkmark & \checkmark & \checkmark & \checkmark & \checkmark & \checkmark \\
    Family & \checkmark & \checkmark & \checkmark & \checkmark & \checkmark & \checkmark \\
    Lifestyle & \checkmark & \checkmark & \checkmark & \checkmark & \checkmark & \checkmark \\
    Subculture & \checkmark & \checkmark & \checkmark & \checkmark & \checkmark & \\
    Fandom & \checkmark & \checkmark & \checkmark & \checkmark & \checkmark & \\
    \bottomrule
  \end{tabular}
\end{table*}

I would be remiss not to raise social identities and potential identities less explored or not explored at all. Moreover, these larger categories can obscure important subgroups. Neurodiversity, for instance, may be distinguished from disability as a whole~\cite{Seaborn_2023weird,Miranda_2023}. Body characteristics, such as size, weight, and proportions~\cite{Seaborn_2023weird}, can be descriptors or signpost identity, as with the fat positivity movement and little people as examples~\cite{Seaborn2023diversity,Miranda_2023}. Religion can intersect with and be categorized under ideology, which covers not only the divine but other spiritual, cultural, political, and other value systems. We must also not forget religious categories (albeit not religions per se) like atheist and agnostic (which are not mutually exclusive, either)~\cite{Schiavone_2017athei}. Social identities like alma mater can cross categories, such as mixing professional and personal spheres. Relational identities like student-mentors and senior-junior may operate at individual and group levels. Hobbyist identities run the gamut, from maker communities to the gaming world. People can identify as activists of all stripes. Identities can be niche and dispersed, such as those who identify as furry, an identity that may be hidden ``in real life'' but openly broadcast online~\cite{Reysen_2015}. While I cannot cover all identities here, I hope the reader can get a sense of the scope and spread of identities that may be considered for HAI contexts.


\section{The Uncanny Killjoy: A Clarion Call}
\label{sec:uncanny}

Now the reader has a good sense of what SIT is all about and how it can---and does---impact the design and deployment of our artificial agent creations. Here, I raise the \emph{ethical implications} of this knowledge. We are always engaging in social identity work with our agents, whether we realize it or not, and whether we wish to or not (\textax{Axiom 1}). As long as agents are socially situated and made in our image (or in the image of other social agents, including non-human animals and social machines from the public imaginary) we must grapple with the potentials and consequences of social identity. Even though the doing of social identity is largely on the human side, machines are created by people and used by people other than the creators. This results in a power imbalance between creators and the rest. 

In response, I suggest an ethical orientation that I call the \emph{uncanny killjoy}. This combines the ``uncanny'' from Uncanny Valley theory~\cite{mori1970bukimi,Mori_2012} as ``eerie'' with ``killjoy,'' meaning to spoil the fun. 
While a bit cheeky, the idea is that we must be extraordinarily cautious, if not embrace the eeriness of the machine (\textax{Axiom 0}). Being an uncanny killjoy is \emph{noetic action} in practice: an act of centring and preemptively reflecting on the potential ethical impacts of our artificial and agentic  creations~\cite{seaborn2025insidious}. Whether on-board, unaware, or unconvinced, we can play devil's advocate by taking on the mantle of the uncanny killjoy, even for a sprint in the design process. Ultimately, my SIA-grounded thesis is that obscuring the uncanny origins of our robotic creations is a decision that must be made with care. 
The uncanny killjoy assists by way of provocation, asserting that the machine must always be laid bare. At minimum, any steps towards anthropomorphism must be taken with extreme caution and solid reasoning.

I now outline several implications for the creator and user sides. A deeper take can be found in \citet{seaborn2024botsagainstbias}.

\subsection{Social Identities are an Ethical Matter}
\label{sec:frameworks}
We must treat social identities derived from the human world with great care. The consequences from SIA (\autoref{sec:consequences}) need to be considered in light of power relationships. For instance, group cohesiveness may be disrupted by a robot that is unable to consistently recognize people with dark skin as group members, echoing the computer vision mishaps identified by \citet{buolamwini2018shades}. This may be a form of identity rejection (of dark-skinned members) or favouritism (of consistently recognized light-skinned members). Distrust in all members is sure to result. Given that social groups are influential, such treatment by an in-group robot may lead to a polarized response towards the robot \emph{or} the person and conformity behaviour. Discriminatory expulsion of the dark-skinned member could result. We may not be able to predict who or what would be most influential.

Avoiding such problems requires an ethical orientation. We must draw from the wealth of academic and activist work on feminism, anti-racism, anti-colonialism, anti-ageism, disability rights, and more. We should not isolate social identities, despite my example above. Intersectionality theory and the matrix of domination explain how power and injustice cascades when multiple social identities come together~\cite{hill2000black,crenshaw2013demarginalizing}. In the foundational example by \citet{crenshaw2013demarginalizing}, Black American women experience greater and different forms of discrimination than their Black men and white women counterparts. Work contextualizing these frameworks for HAI and adjacent domains has begun~\cite{Ostrowski_2022} and should be referenced as a guide when making decisions about agent social identities for specific contexts and user groups, e.g., feminist HCI~\cite{Bardzell_2010}, feminist HRI~\cite{Winkle_2023}, postcolonial computing~\cite{Irani_2010}, domains of expert power~\cite{ Williams_2024}, intersectional HCI~\cite{Schlesinger_2017}, disability ethics~\cite{Allen_2024}, and critical race theory for HCI~\cite{Ogbonnaya_Ogburu_2020}, to name a few.
    
    
\subsection{Superficial and Unnecessary Social Identity Detection is Unethical}
Personalization and identification based on user faces, bodies, and voices has a long history within computer vision, robotics, and increasing AI systems of all kinds. Through these technologies, machines can partake in superficial and dangerous social identity activities~\cite{Keyes_2018,benjamin2019race,Williams_2023}. The reasons may be benign and incidental, like the computer vision mishap of poorly trained camera AI~\cite{buolamwini2018shades}, but the result is the same: the agent literally does not ``see'' Black women. 
\citet{Williams_2023} outlines the situation for HRI, and while concentrating on race, these ideas transcend social identity: social identities are difficult to operationalize and may even be inappropriate for operationalization; those responsible may have limited or harmful conceptual models of any given social identity; social identities are dynamic, and the system would have to react in kind; and the agent (or its overseers) may use this information in harmful ways, whether intentionally or not. A social identity approach raises the additional problems of influence and conformity. A prototypical agent may, for instance, influence humans to socially categorize others in line with its superficial and harmful classifications, or worse: incite action that is injurious. We must question the what, how, and why of simple social identity detection based on complex human social identities, and aim for methods more authentic. 

\subsection{Social Identity Cues Will Be Found, No Matter How Subtle}
Agents can be designed or primed with social identities that people pick up and respond to. We cannot help ourselves: we will ``read'' these patterns, no matter how fine-drawn. The cue can be seemingly benign, such as a name~\cite{Barfield_2024,Kuchenbrandt_2013} or pronoun~\cite{Seaborn_2022pepper,Fujii_2024}. No matter; as per the CASA~\cite{nass1995can,nass2005wired,lee2000can} and related models, we tend to react unthinkingly to social identity cues on the first instance, most of the time. 
A classic example is  \citet{Kuchenbrandt_2013}, who replicated the 
foundational results in \citet{Tajfel_1971overunder} for the NAO robot. 

Echoing \citet{Vanman_2019}, I suggest that we maintain a certain level of uncanniness and artificiality in our agents---for our own good. Some may worry that this will limit the desired user experiences with these machines. Yet, a wealth of research has shown that it is more a matter of psychological attachment than physical realism. The mechanical Sony AIBO dog robot, for instance, has lived well past its life cycle thanks to the concerted efforts of loving ``pet owners,'' to the extent that unfixable AIBOs may be given Buddhist burial rites~\cite{Knox_2018}. SIA and specifically SIT would explain this as positive group distinctiveness: the AIBO has become a valued member of the family, despite its uncanniness.

\subsection{Social Identity Theories are Biased}

Most scientific knowledge is English, Western-based, and Anglocentric~\cite{Henrich_2010,Levisen_2019}. This is true for HCI~\cite{Linxen_2021weird}, HRI~\cite{Seaborn_2023weird}, and adjacent human participant domains~\cite{Henrich_2010}. The ideas mapped out here may not be appropriate to all cultures or times. One way forward is to explore culturally-sensitive~\cite{Seaborn_2024cultural} alternative mappings. Buddhism, for instance, undergirds key work in robotics, like the Uncanny Valley theory, developed through creator Masahiro Mori's study of and inspiration from Mahayana Buddhism and Zen Buddhism~\cite{kimura2018masahiro}. Buddhist philosophies and worldviews may be deployed in identity work within HRI contexts~\cite{Miranda_2024identity,white20259}, but this must be done thoughtfully and with care, ideally by working with experts on Buddhism and ensuring citational justice~\cite{Collective2021} to core Buddhist texts and existing Buddhist (and often non-Western) HRI scholarship.
Social identity approaches are also biased to the superordinate identity of ``human,'' i.e., anthropocentric. 
For the sake of future self-aware machines, I also include a plea to my fellow humans: we are a dominant species, warmongers and executors of much evil~\cite{vetlesen2005evil}. Science fiction is peppered with examples of our fear incarnate: that a machine species will arise and take over. Scholarship has explored these notions with respect to non-human animals and non-animals~\cite{Blount_Hill_2021}. If we dare to bring the machine to life, let it live with us in peace and harmony.

\subsection{Being the Uncanny Killjoy: Practical Examples}

Awareness is crucial, but being an uncanny killjoy requires action. Yet, the wealth of agent and human embodiments in existence and imaginable are so broad that a complete list of practical examples is difficult to articulate and likely to change. Tailored frameworks for specific embodiments---humanoid robots, zoomorphic robots, virtual agents, and more---that include room for multi-embodiments~\cite{branksky2024bodies} and disciplinary perspectives, such as communication studies~\cite{Urakami_2023nonverbal} will be ideal. I present some starting points for expansion in future work:

\begin{itemize}
    \item \emph{Create critical checklists:} The uncanny killjoy may offload the cognitive process of self-checks to a checklist. Established ethical and critical frameworks, such as those referenced in \autoref{sec:frameworks}, could be formalized as checklists that creators, researchers, and practitioners could use to ensure that the agent, the users, and the interaction between them has been ethically considered.
    
    \item \emph{Critique detection mechanisms:} All user detection mechanisms---and their constituents, from the data to the algorithm to the testing means---should be critically appraised with respect to social identities. An uncanny killjoy may even avoid the use of such mechanisms and design agents that could, for example, simply ask people how they identify. People should also have the autonomy to decline, as certain knowledge could be illegal to collect (as race data can be) or dangerous to hold (given that machines can be hacked or tricked into disclosures, just like people)~\cite{Olteanu_2019}. 

    \item \emph{Cue the uncanny:} The uncanny killjoy first presents the agent as artificial. This is not an immutable category. Rather, cues to artificiality can be designed into the agent, however humanlike it may be otherwise. For example, nonverbal cues like mechanical glitches in a robot's voice, verbal cues that periodically reinforce the superordinate identity (``I'm a robot, after all''), and uncanny embodiments like metal plating on robot ``skin'' could suffice to intimate that the agent is a machine.
\end{itemize}



\section{Conclusion}

Humans are social creatures. We who craft and deploy humanlike machines are no stranger to the notion that leveraging anthropomorphic cues and styles of social interaction from the human world can be useful, and even unavoidable. Here, I have provided a theoretical framework less explored within HAI circles but well-established within the social sciences. I have provided examples---actual and imagined---of how this theory and its various sub-theories apply to agents and HAI scenarios. Largely, I have centred this around involvement of a human actor; however, if and when artificial agents attain the ability to fully participate in social identity activities, this framing may need to shift. I am personally invested in a match between the droids of \emph{Star Wars} and Data from \emph{Star Trek}. In ending, I have issued a clarion call for awareness and restraint: the closer we bring our machine creations to peak humanlikeness, the greater the danger of falling prey to mechanical deceits. We have a grave responsibility to design and deploy agents with awareness and a sense of ethics.

The HRI and broader HAI and HCI communities of practice continue to invest in the question of identity as a factor of agent design. The HRI workshop series opened with \citet{lee2021roboidenws} raising the fundamental question of robot identity and proposing ethical and philosophical perspectives engagement with the notions of dynamic embodiment and multiple bodies. \citet{laban2022roboident2} focused on synthetic speech as a crucial mode of identity and a prompt for questions on the degree of agent artificiality, hinting at an uncanny killjoy perspective. For the third iteration, \citet{Khot2024roboidenws} wove both perspectives into the reality of our social environment comprised of diverse users, highlighting the need for identity inclusion and anti-bias work. The three themes of the workshop---altogether, personal and social identity, designed and deployed in the shared world---strikingly parallel my efforts here. This primer may guide the next phase of ``robo-identity'' work and help deepen the discourse on SIA in HRI from a theoretical lens.

I will end with a few reflective prompts, starting with: What mixture of utopia with uber-humanlike agents and dystopia with a lack of control and susceptibility do we wish to create? The dream of many ``parents'' is to replicate ourselves in our ``offspring.'' From the conversationally fluent ChatGPT to doppelgängers like the Geminoid series of androids~\cite{nishio2007geminoid}, exceptionally anthropomorphic agents continue to front advances in agentic technologies and capture our collective imaginations. When the magic wears off, where do we wish to be? Do we wish for a world wherein artificial agents that reach the point of human-level SIT ability necessarily influence our own identities? When autonomy, sufficiently advanced artificial intelligence, and SIT meet, humans may no longer be in control. What artificial wilderness do we wish to traverse? These are but of a few questions haunting my imagination, and I hope the reader's, too.

\begin{acks}
Thank you to Peter Pennefather for pre-reviewing this paper, as well as lively conversations on the topic of artificial intelligence and social identity. Thank you to the reviewers of the original workshop paper for their excellent feedback.
\end{acks}

\bibliographystyle{ACM-Reference-Format}
\bibliography{main}










\end{document}